\documentclass[twocolumn,prl,showpacs,amssymb,superscriptaddress]{revtex4-1}

\usepackage{graphicx}
\usepackage{amsmath}
\usepackage{amsfonts}
\usepackage{amsthm}
\usepackage{amssymb}
\usepackage{amsbsy}
\usepackage{wasysym}
\usepackage{bbm}
\usepackage{bm}
\usepackage{mathrsfs}
\usepackage{color}
\usepackage{hyperref}
\usepackage{braket}
\usepackage{ulem}
\setcitestyle{super}

\newcommand*{\citen}[1]{%
  \begingroup
    \romannumeral-`\x 
    \setcitestyle{numbers}%
    \cite{#1}%
  \endgroup   
}

\date{\today}
\begin{document}
\title{Subdiffusive front scaling in interacting integrable models}

\author{Vir B. Bulchandani}
\affiliation{Department of Physics, University of California, Berkeley, Berkeley CA 94720, USA}

\author{Christoph Karrasch}
\affiliation{Dahlem Center for Complex Quantum Systems and Fachbereich Physik, Freie Universit{\"a}t Berlin, 14195 Berlin, Germany}
\affiliation{Technische Universit{\"a}t Braunschweig, Institut f{\"u}r Mathematische 
Physik, Mendelssohnstra{\ss}e 3, 38106 Braunschweig, Germany}

\begin{abstract}
We show that any interacting integrable model possesses a class of initial states for which the leading corrections to ballistic transport are subdiffusive rather than diffusive. These initial states are natural to realize experimentally and include the domain wall initial condition that has been the object of much recent scrutiny. Upon performing numerical matrix product state simulations in the spin-$1/2$ XXZ chain, we find that such states can exhibit subdiffusive $t^{1/3}$ scaling of fronts of spin, energy and entanglement entropy across the entire range of anisotropies. This demonstrates that Tracy-Widom scaling is not incompatible with model interactions, as was previously believed.
\end{abstract}

\maketitle

\paragraph{Introduction.}

The typical relaxation dynamics of conserved quantities such as energy and particle number in classical many-body systems has been understood for well over a century, and is described to a great degree of accuracy by phenomenological ``laws of diffusion'', such as Fick's law and Fourier's law. At the same time, attempts to derive these laws from a microscopic model of deterministic Hamiltonian evolution are still in their infancy, owing to the tremendous technical difficulties involved in such a task\cite{Bonetto2000}. These difficulties are perhaps related to a growing understanding that classical transport in low dimensions is frequently \textit{anomalous}, in the sense of being characterized by quantities that exhibit $t^{1/3}$ scaling with time\cite{VanBeijeren2012, Spohn2014} corresponding to the KPZ universality class of dynamics, rather than ordinary, microscopic Brownian motion that would give rise to diffusive $t^{1/2}$ scaling.  

These recent advances in the theory of low-dimensional classical transport have been complemented by the development of a hydrodynamic theory of time-evolution in quantum integrable models\cite{Castro-Alvaredo2016,Bertini2016}, which usually goes by the name of ``generalized hydrodynamics'' (GHD). Quantum integrable models include experimentally realizable examples like the Lieb-Liniger gas of delta interacting bosons in one spatial dimension and the spin-$1/2$ Heisenberg chain. At first, generalized hydrodynamics was limited to describing ballistic transport in such models, for which it has so far yielded impressive agreement with numerical simulations\cite{Castro-Alvaredo2016,Bertini2016,Ilievski2017,BVKM2018,BVKM2017,Doyon2017,Ilievski20172,Piroli2017,Alba2017,Collura2018,Urichuk2018}.
However, the diffusive corrections to the ``Bethe-Boltzmann'' equation underlying the hydrodynamic approach were recently derived\cite{DeNardis2018,Gopalakrishnan2018}, and have raised the novel possibility of using generalized hydrodynamic techniques to analyze sub-leading corrections to ballistic transport.

In the present work, we study time-evolution from a class of initial states in the spin-$1/2$ XXZ chain with anisotropy $\Delta$, for which these recent results predict that the diffusive corrections to the Bethe-Boltzmann equation ought to vanish. Upon performing numerical simulations using the real-time density matrix renormalization group\cite{White2004,Schollwoeck2011,Karrasch2014}, we find that these states exhibit subdiffusive $t^{1/3}$ scaling of fronts of spin, energy and entanglement entropy across the entire range of anisotropies, which we interpret to be a consequence of third-order derivative terms\cite{Fagotti2017} in the hydrodynamics of the propagating front. The class of states we discuss includes a domain-wall initial state that was found to support superdiffusive transport at the isotropic point\cite{Gobert2005,Ljubotina2017} $\Delta=1$; it was subsequently argued that this effect might be a transient deviation from diffusive transport\cite{Misguich2017}. The fact that the leading corrections to ballistic transport are diffusive away from the isotropic point is supported by studies of domain-wall initial states in the gapless phase, $|\Delta|<1$, including a hydrodynamic argument\cite{Collura2018} that fronts of spin scale with time as $t^{1/2}$, together with an analytical study of return probabilities in the six-vertex model that also found the $t^{1/2}$ scaling characteristic of diffusion\cite{Stephan2017}. 

Before presenting our results in detail, we briefly summarize how they relate to these earlier analyses of time evolution from domain-wall initial states. For time evolution from domain walls with $|\Delta|<1$, we find that ballistic fronts of spin and energy propagate at a light-cone speed $v_* = 1$, rather than the value $v_* = \sqrt{1-\Delta^2}$ found in previous works\cite{Collura2018,Stephan2017}. Spreading of observables in the ``forbidden'' region $\sqrt{1-\Delta^2} < x/t < 1$ was explicitly noted in Ref. \citen{Stephan2017} (and indeed earlier\cite{Gobert2005}), but characterized as a transient effect. Here, we argue that this discrepancy with theory is due to a specific choice of ``coarse-graining'' in earlier works, corresponding to the ansatz Eq. \eqref{initcond} for the hydrodynamic initial state, which omits a finite energy $\delta E = - J\Delta/2$ at the domain wall itself. In the ballistic scaling limit as $t \to \infty$, this sub-extensive contribution to the dynamics vanishes, and the usual GHD prediction\cite{Castro-Alvaredo2016,Bertini2016} is presumably exact. However, in the context of GHD on finite time and length scales\cite{Doyon2016,BVKM2017,BVKM2018,Doyon2017}, the inhomogeneity cannot be neglected, and generates quasiparticles that propagate ballistically through the system for all $t < \infty$, with a light-cone speed $v_*=1$. Upon performing a scaling analysis of tDMRG data at this physical light-cone edge, we find $t^{1/3}$ behaviour rather than diffusive $t^{1/2}$ scaling; an example is shown in Fig. \ref{Fig2}. 

\begin{figure}[t]
\includegraphics[width=\linewidth]{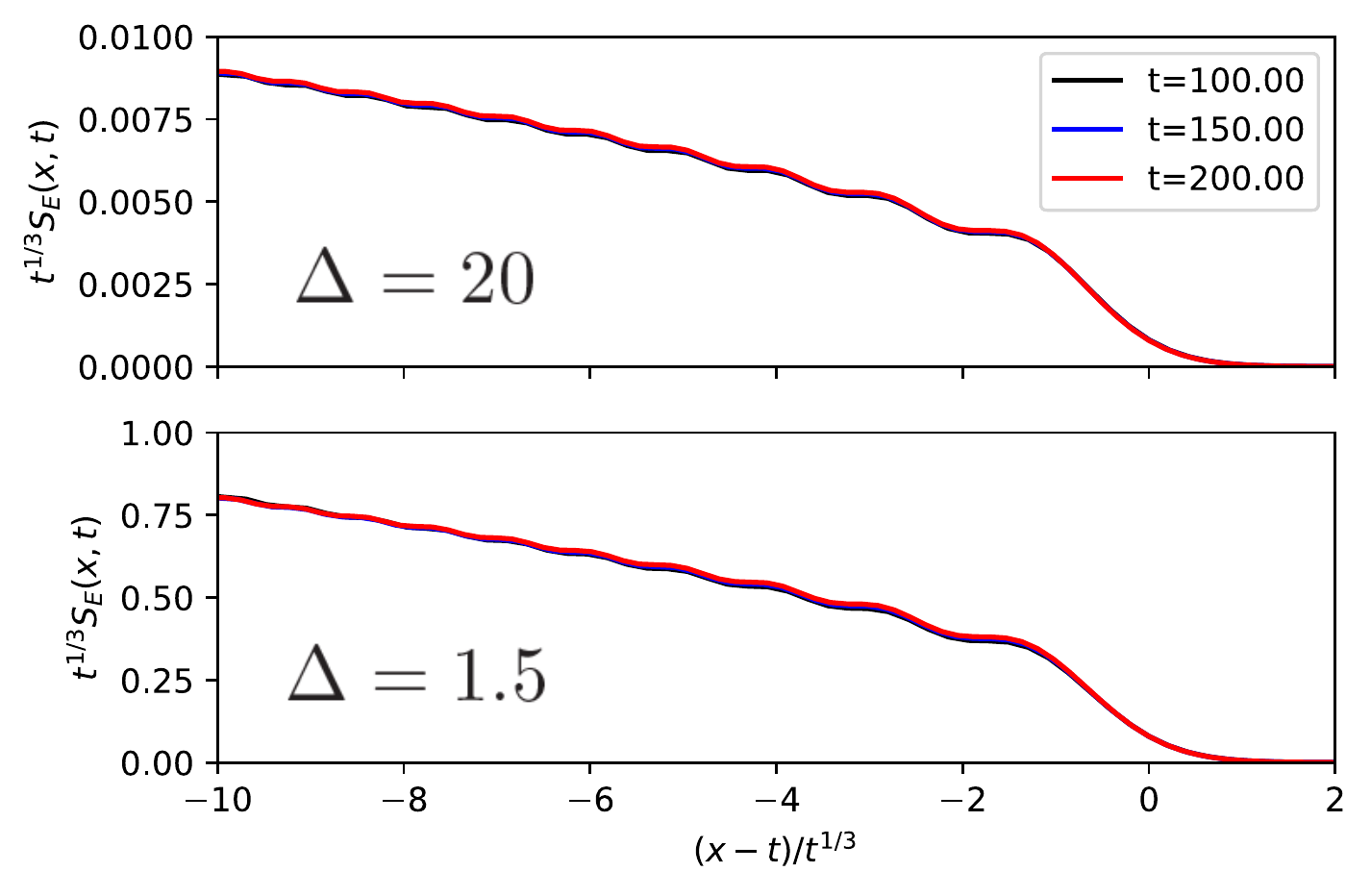}
\caption{Scaling of fronts of the single-cut entanglement entropy in tDMRG data for time-evolution from domain wall initial conditions in the gapped phase of the XXZ chain, with example anisotropies $\Delta=1.5$ and $\Delta=20$. The $t^{1/3}$ scaling collapse, together with the ``staircase feature'' characteristic of the Airy kernel, are clear.}
\label{Fig1}
\end{figure}

\begin{figure*}[t]
\includegraphics[width=\linewidth,clip]{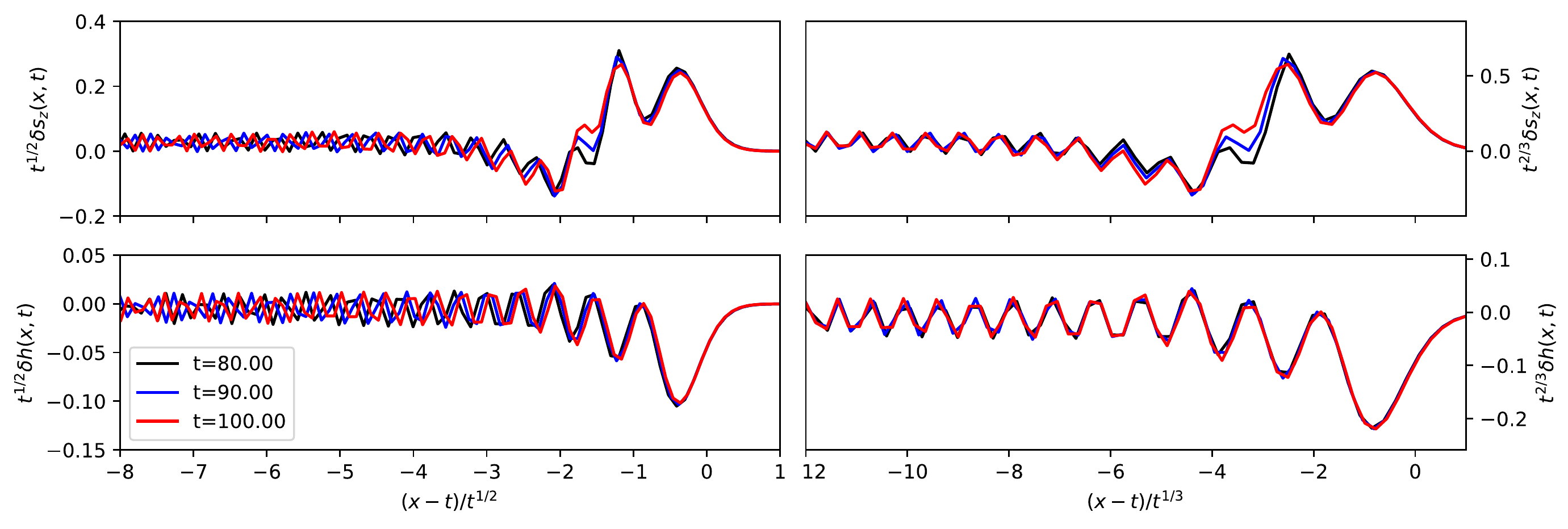}
\caption{Diffusive (\textit{left}) versus subdiffusive (\textit{right}) rescaling of fronts of spin (\textit{top}) and energy (\textit{bottom}), obtained from tDMRG predictions for time-evolution from domain wall initial conditions in the gapless phase of the XXZ chain, with anisotropy $\Delta=0.5$. The improvement in scaling collapse for the right-hand figures is marked.}
\label{Fig2}
\end{figure*}

We now turn to anisotropies $\Delta \geq 1$, for which conventional wisdom predicts no ballistic propagation of observables at all. Once again, this is based on the ``two-reservoir'' hydrodynamic ansatz Eq. \eqref{initcond}, which is susceptible to sub-leading corrections at finite times. Indeed, in the absence of extensive contributions to ballistic transport, the lattice-scale inhomogeneity at the domain wall may dominate the finite-time dynamics. From numerical tDMRG simulations, we observe that the inhomogeneity generates ballistically spreading fronts of energy, spin and entanglement entropy. As in the gapless phase, the light-cone speed for the associated quasiparticles is $v_* = 1$, and the spreading of the quasiparticle front is subdiffusive in time, scaling as $t^{1/3}$. See Fig. \ref{Fig1} for examples with $\Delta>1$, or the Appendix for examples with $\Delta=1$.

\paragraph{Hydrodynamics of integrable models.}

Consider a generic quantum integrable model, whose equilibrium states may be characterized in terms of quasiparticle distribution functions $\rho_{n,k}$, with $n \in \mathbb{N}$ a discrete quasiparticle index and $k \in \mathbb{R}$ a continuous rapidity variable. There is now a substantial body of numerical evidence\cite{Castro-Alvaredo2016,Bertini2016,Ilievski2017,BVKM2018,BVKM2017,Doyon2017,Ilievski20172,Piroli2017,Alba2017,Collura2018,Urichuk2018} that the ballistic part of time-evolution in such models, from smooth, locally equilibrated initial conditions, i.e. those that can be modelled by smoothly varying distribution functions $\rho_{n,k}(x)$, may be captured by the system of Boltzmann-type equations\cite{Castro-Alvaredo2016,Bertini2016}
\begin{equation}
\label{BB}
\partial_t \rho_{n,k} + \partial_x (\rho_{n,k} v_{n,k}[\rho]) = 0,
\end{equation}
where the local quasiparticle velocities $v_{n,k}(x,t)$ at each space-time point are fixed in terms of the full set of local distribution functions $\{\rho_{n',k'}(x,t) : n' \in \mathbb{N}, k' \in \mathbb{R}\}$ via thermodynamic Bethe ansatz. As it stands, the system \eqref{BB} conserves the local Yang-Yang entropy density at each point (as follows from its time-reversal invariance) and so cannot capture diffusive effects. However, the leading diffusive correction to this ballistic hydrodynamics was recently derived by taking two-body scattering processes into account\cite{DeNardis2018} and upon including this correction, the entropy-conserving system \eqref{BB} is replaced by a dissipative system of equations, of the form
\begin{equation}
\label{BBD}
\partial_t \rho_{n,k} + \partial_x (\rho_{n,k} v_{n,k}[\rho]) = \partial_x (\mathcal{D}_{n,k}[\partial_x \rho]).
\end{equation}
where the ``diffusion operator'' $\mathcal{D}_{n,k}$ acts on $\partial_x \rho$ as a linear integral kernel. Meanwhile, a kinetic theory argument based on the propagation of a tagged soliton through a fluctuating medium\cite{Gopalakrishnan2018} predicts the linear diffusion equation
\begin{equation}
\label{BBD2}
\partial_t \delta\theta_{n,k}+v_{n,k}[\theta]\partial_x \delta\theta_{n,k} = D_{n,k}[\theta]\partial_{xx}\delta \theta_{n,k}
\end{equation}
for small perturbations $\delta \theta_{n,k}(x,t)$ of a locally equilibrated background with local Fermi factors $\{\theta_{n',k'}(x,t): n' \in \mathbb{N}, k' \in \mathbb{R}\}$. This turns out to coincide with the diagonal, linear-response component of the full transport equation Eq. \eqref{BBD}, as might be expected from its derivation\footnote{To be specific, the derivation of Eq. \eqref{BBD2} considers perturbations only in the channel with quasiparticle index $n$ and rapidity $k$, and neglects any back-reaction effects that would generate off-diagonal couplings to indices $n' \neq n$ or rapidities $k'\neq k$. Such back-reaction effects are taken into account by the full transport equation Eq. \eqref{BBD}, but the conclusions of the present work appear to be insensitive to these corrections.}. In the present work, our arguments are based on the qualitative picture leading to Eq. \eqref{BBD2} but our conclusions are also consistent with the full transport equation, Eq. \eqref{BBD}.

\paragraph{Diffusive vs subdiffusive corrections to hydrodynamics.}

For non-interacting integrable models, in the sense of Ref. \citen{Spohn2018}, the possibility of subdiffusive $t^{1/3}$ scaling of ballistic fronts is well-established by now\cite{Hunyadi2004,Platini2005,Eisler2013,Kormos2017}, and was recently given a new interpretation as an effect in ``third-order hydrodynamics''\cite{Fagotti2017}\footnote{Here, ``third-order'' refers to the order of the derivative terms involved, by contrast with the usual terminology, according to which these are second-order corrections to zeroth-order, or Euler, hydrodynamics.}, which is characterized by the absence of diffusive terms. By contrast, a recently discovered link between linear-response diffusive corrections to the Bethe-Boltzmann equation and local density fluctuations\cite{Gopalakrishnan2018} indicates that for locally equilibrated states of interacting integrable models, the generic scaling of operator fronts goes as $t^{1/2}$, rather than $t^{1/3}$. The key insight is that the density fluctuations giving rise to microscopic diffusion are controlled by fluctuations $\delta \theta_{n,k}$ in the local Fermi factors, which satisfy\cite{Fendley1996}
\begin{equation}
\label{fluct}
\langle \delta \theta_{n,k} \delta \theta_{n',k'} \rangle \propto \delta_{n,n'}\delta(k-k')\theta_{n,k}(1-\theta_{n,k})
\end{equation}
on a given interval. In order for diffusive corrections to vanish, the Fermi factors $\theta_{n,k}$ must vanish for all $n$ and $k$ at every space-time point. Thus, as claimed in Ref. \citen{Gopalakrishnan2018}, a generic local equilibrium state of an interacting integrable model will exhibit diffusive corrections to ballistic dynamics, and consequently $t^{1/2}$ scaling of operator fronts. The observation that we wish to make in the present work is that the ``exceptional case'' for which there is vanishing entropy production in the majority of the system, or equivalently, for which $\theta_{n,k}$ is $0$ or $1$ almost everywhere, is physically rather natural. For example, in the context of spin-$1/2$ XXZ chains, this class of states encompasses any initial condition that consists of macroscopically large ferromagnetic domains with spin alignment along the $z$-axis, which are simple to realize experimentally in spin-chain compounds\cite{Sologubenko2007}. Similar zero-entropy initial states were considered in the context of Lieb-Liniger Bose gases\cite{Doyon2017}. On the basis of the formula Eq. \eqref{fluct} and recent theoretical results on third-order derivative terms in GHD\cite{Fagotti2017}, we conjecture that for any integrable lattice model with nearest-neighbour interactions, initial states that are ``pseudo-vacua'' in bulk will support $t^{1/3}$ corrections to ballistic dynamics, provided that the bare dispersion of the fastest quasiparticle satisfies the technical conditions discussed in Ref. \citen{Fagotti2017}. The single magnon excitation in the spin-$1/2$ XXZ chain satisfies these conditions\footnote{This is immedate from its cosine dispersion and full pseudomomentum range. In fact, for $\Delta \geq 1$ it is additionally true for \emph{all} string excitations, for the same reason\cite{Takahashi}.} for all values of the anisotropy $\Delta$, allowing for a direct test of our predictions against matrix product state numerical simulations.

Upon performing a scaling analysis of tDMRG data for spin, energy and single-cut entanglement entropy in two classes of such initial states in the spin-$1/2$ XXZ chain, namely i) domain wall initial conditions, which have come under recent scrutiny\cite{Bertini2016,Ljubotina2017, Collura2018, Stephan2017,Misguich2017} and ii) initial conditions consisting of a single flipped down spin in a sea of up spins, we find evidence for $t^{1/3}$ scaling of fronts across the entire anisotropy range of the XXZ chain. In more detail, for the initial condition (i) with $\Delta \geq 1$, we find that the single-cut entanglement entropy at a point $x$ exhibits front scaling
\begin{equation}
\label{EEscaling}
S_E(x,t) \sim t^{-1/3}f\left(\frac{x-t}{t^{1/3}}\right),
\end{equation}
consistent with the usual Airy kernel\cite{Hunyadi2004,Platini2005,Eisler2013} (see Fig. \ref{Fig1}), while for both initial conditions (i) and (ii) and all anisotropies $\Delta$, fronts of spin and energy exhibit the scaling form (see Fig. \ref{Fig2} and Appendix) 
\begin{equation}
\delta \langle \mathcal{O} \rangle (x,t) \sim t^{-2/3}g\left(\frac{x-t}{t^{1/3}}\right),
\end{equation}
consistent with the derivative of the Airy kernel, as was found to capture front-scaling in the critical transverse-field Ising chain\cite{Kormos2017}. A detailed analysis of these scaling functions is beyond the scope of the present work, though some analytical results for the spin-flip initial condition are summarized in the Appendix.

\paragraph{Initial states supporting subdiffusion of fronts in the XXZ model.}

We now discuss the initial conditions (i) and (ii) in more detail. The Hamiltonian under consideration is the spin-$1/2$ XXZ chain with anisotropy $\Delta$, namely
\begin{equation}
H = J\sum_{i=1}^{L} S_i^x S_{i+1}^x + S_i^y S^y_{i+1}+ \Delta S_i^z S_{i+1}^z.
\end{equation}
The first class of initial states we consider has the form
\begin{equation}
\label{initialdw}
|\psi\rangle =  \ket{\uparrow}^{\otimes L/2} \otimes \ket{\downarrow}^{\otimes L/2}.
\end{equation}
This initial condition has been studied frequently in recent works, both at the isotropic point\cite{Ljubotina2017,Misguich2017} $\Delta=1$ and in the gapless phase\cite{Bertini2016,Collura2018,Stephan2017} of the XXZ model. 

At this point, a remark is in order on the standard hydrodynamic model for time evolution from ``two-reservoir'' initial conditions, corresponding to initial density matrices of the form $\rho = \rho^L \otimes \rho^R$, where $\rho^{L/R}$ denote GGE density matrices with associated Fermi factors $\theta_{n,k}^{L/R}$. By now, standard practice is to model the time evolution of $\rho$ by evolving the hydrodynamic initial state
\begin{equation}
\label{initcond}
\theta_{n,k}(x,0) = \begin{cases} \theta_{n,k}^L(x) & x \leq 0 \\ \theta_{n,k}^R(x) & x > 0\end{cases}
\end{equation}
under ballistic GHD, to yield a hydrodynamic profile $\theta_{n,k}(x,t)$ that is expected to be exact in the ballistic scaling limit as $t \to \infty$. Even if one assumes that the hypothesis of local equilibrium is valid at all times, this model of time dynamics is susceptible to two types of hydrodynamic corrections\cite{Doyon2016,Piroli2017}. The first type are higher-order derivative terms in the hydrodynamic equations\cite{Fagotti2017,DeNardis2018,Gopalakrishnan2018}. These capture physical effects like diffusion (which appears at second order) or lattice corrections (at third order and above). The second type of hydrodynamic corrections arise at the level of initial conditions and are generated by sub-extensive numbers of quasiparticles, that may nevertheless dominate dynamics on finite length and time scales. The effects studied in the present work arise from the interplay of both types of correction. It seems worth emphasizing that for two-reservoir initial states $\rho^L \otimes \rho^R$, there is always a finite-length correction to the standard initial condition Eq. \eqref{initcond}, due to the inhomogeneity at $x=0$. Often, this correction is negligibly small, but for certain states like domain walls with $|\Delta| > 1$, it can yield the dominant contribution to time dynamics.

To illustrate how one might describe the finite-length correction to Eq. \eqref{initcond} in practice, it is instructive to consider a rather simpler initial state, namely the localized spin-flip
\begin{equation}
\label{initialflip}
|\psi\rangle =  \ket{\uparrow}^{\otimes L/2-1} \otimes  \ket{\downarrow} \otimes \ket{\uparrow}^{\otimes L/2}.
\end{equation}
For this state, a hydrodynamic coarse-graining procedure that neglects sub-extensive corrections, as is implict in Eq. \eqref{initcond}, would predict no time evolution at all. However, the exact time evolution of a localized magnon can be obtained directly\cite{Antal1999,Hunyadi2004} (see Appendix). In particular, its large-scale dynamics is given by evolving the initial distribution $\rho_{1,k}(x) = \delta(x)$ of magnons under the hydrodynamic equation derived in Ref. \citen{Fagotti2017}, which yields ballistic spreading with light-cone speed $v_*=1$ and $t^{1/3}$ broadening of fronts.

Similarly, we postulate that for domain wall initial conditions, there is always a finite-length correction to the hydrodynamic initial condition Eq. \eqref{initcond} in the single magnon sector, of the form $\rho_{1,k}(x) = C(\Delta)\delta(x)$, for some model dependent weight $C(\Delta)$. This postulate, which can be proved\cite{BVKS} as $\Delta \to \infty$, and is justified on physical grounds by the uncertainty principle (in conjunction with the spatial localization of the initial inhomogeneity to $x=0$), yields the following predictions for the time dynamics of the state Eq. \eqref{initialdw}.

First, it implies that the fastest quasiparticles travel at the magnon light-cone speed, $v_* =1$, which is consistent with our observations and has been noted previously\cite{Gobert2005,Misguich2017,Stephan2017}. Next, it implies that these quasiparticles travel in a bulk pseudovacuum, experience no microscopic diffusion and can be described by the hydrodynamic equation derived in Ref. \citen{Fagotti2017} (see discussion above). It follows that the degrees of freedom at the edge of the front are free-particle like, and that the particle density profile there coincides with the edge of the domain wall front in the XX model, up to rescaling by the overall factor $C(\Delta)$. From this, it is immediate that broadening of the front scales as $t^{1/3}$ in time, and that the position of the fastest quasiparticle is described by Tracy-Widom statistics\cite{Eisler2013}. We now directly verify the prediction of $t^{1/3}$ scaling against microscopic tDMRG simulations.

\paragraph{$\Delta \geq 1$: domain-wall initial conditions.}

As discussed above, the localized initial energy density at $x=0$ gives rise to ballistic quasiparticle spreading even in the regime $\Delta \geq 1$. This is particularly clear in the profiles of single-cut entanglement entropy, whose fronts propagate ballistically in time for all values of the anisotropy $\Delta \geq 1$, in a manner consistent with the creation of counter-propagating pairs of magnons at the domain wall itself\footnote{More precisely, for $\Delta \gtrsim 5$, the domain wall appears to generate mainly single-magnon excitations, while as $\Delta \to 1$, higher strings are also excited.} (this has already been noted\cite{Misguich2017} at $\Delta=1$). Entanglement spreading from domain wall initial conditions in the gapped phase of the XXZ chain will be discussed further in related work\cite{BVKS}. Upon plotting the fronts of single-cut entanglement entropy (as was previously done for free fermions\cite{Eisler2013}), we find that entanglement fronts exhibit the $t^{1/3}$ scaling previously thought to be specific to non-interacting systems, together with the ``staircase'' feature characteristic of the Airy kernel\cite{Hunyadi2004,Eisler2013,Kormos2017}. See Fig. \ref{Fig1}.

\paragraph{$|\Delta| < 1$: domain-wall initial conditions.}

For $|\Delta| < 1$, evolution of entanglement from domain-wall initial conditions, as was considered above, exhibits non-ballistic growth in time due to the gaplessness of magnon excitations on each pseudo-vacuum. However, as for $\Delta \geq 1$, there is ballistic transport of spin and energy. To our surprise, the standard scaling analysis of spin fronts\cite{Hunyadi2004,Eisler2013} does not appear clearly to distinguish between diffusive and subdiffusive scaling. We put this ambiguity down to the dominance of ballistic transport and therefore subtract the ``two-reservoir'' hydrodynamic prediction for the steady-state spin density, given at roots of unity $\Delta = \cos{\gamma}$, $\gamma = \pi/\nu$, $\nu \in \{2,3,\ldots\}$, by\cite{Collura2018,Stephan2017}
\begin{align}
\label{hydropred}
s^{hydro}_z(x,t) &= \begin{cases} 0.5 & x<-(\sin{\gamma})t \\ -\frac{1}{2{\gamma}} \sin^{-1}\left(\frac{x}{t}\right) & |x|< (\sin{\gamma})t \\ -0.5 & x > (\sin{\gamma})t \end{cases}
\end{align}
Once this is subtracted from the numerical data, the oscillatory features in the spin front show a marked collapse to $t^{1/3}$ rather than $t^{1/2}$ scaling; see Fig. \ref{Fig2} for an example with $\Delta=0.5$. For the energy fronts, we merely subtract the bulk value $\langle h \rangle = J\Delta/4$ as usual\cite{Hunyadi2004}, and observe the same scaling. 

In both cases, our rescaled horizontal coordinate involves $x-t$, rather than the parameter $x-\sqrt{1-\Delta^2}t$ plotted in previous work\cite{Collura2018}. This is consistent with the physical light-cone speed, $v_* = 1$. As discussed above, the presence of quasiparticles in the asymptotically forbidden region $\sqrt{1-\Delta^2} < x/t < 1$ is due to the lattice-scale inhomogeneity at $x=0$.

\paragraph{Discussion.}

The above arguments indicate that the ``exceptional'' case of vanishing entropy production in locally equilibrated states of quantum integrable models\cite{Gopalakrishnan2018} in fact includes a class of states that arise quite naturally in practice, since any initial condition that gives rise to a dilute gas of quasiparticle excitations propagating through a bulk pseudo-vacuum will lack the local density fluctuations that generate diffusion. The absence of bulk entropy production in these states can also be seen from the vanishing of the diffusion kernel\cite{DeNardis2018} in the full transport equation, Eq. \eqref{BBD}. Experimental realizations of such physics include the free expansion of a spatially localized initial density of Lieb-Liniger gas and, perhaps more surprisingly, the example of time-evolution from ferromagnetic domain walls in spin-$1/2$ XXZ chains that was discussed in detail above.

Furthermore, our analysis identifies an important class of sub-leading corrections to the usual hydrodynamic description, Eq. \eqref{initcond}, of ``two-reservoir'' type initial conditions. For example, we expect such corrections to be present for the initial states $\rho = (\mathbbm{1}+\mu\sigma^z)^{\otimes L/2} \otimes (\mathbbm{1}-\mu \sigma^z)^{\otimes L/2}$ that were shown to support superdiffusive spin transport at the isotropic point\cite{Znidaric2011,Ljubotina2017} of the spin-$1/2$ XXZ chain (note that in this paper we studied the propagation of \textit{fronts}, not transport properties). Previously, it was assumed that hydrodynamics could say nothing about time evolution from such initial states for $\Delta \geq 1$, because the initial condition Eq. \eqref{initcond} is homogeneous for these states. By contrast, our demonstration that at $\mu=1$, the standard hydrodynamic description of these states needs to be augmented by initial data localized at $x=0$, if it is to yield accurate results at finite times, indicates that the same modification is needed for $\mu<1$. Our results additionally show that third-order effects can generate non-negligible corrections to ballistic dynamics, even in fully interacting integrable models. Whether or not these refinements of generalized hydrodynamics can shed any light not only on the scaling of fronts, but also on the rich variety of transport phenomena observed at the isotropic point\cite{Gobert2005,Znidaric2011,Ljubotina2017,Ljubotina20172,Misguich2017} of the XXZ chain, is an interesting topic for future investigation.

\paragraph{Acknowledgments}
We are grateful to R. Vasseur and T. Scaffidi for collaborations which led to this work, and to X. Cao for helpful discussions on related topics. We thank J. de Nardis, B. Doyon, J. E. Moore, R. Vasseur and M. {\v Z}nidari{\v c} for their comments on the manuscript. We also thank the anonymous referees for valuable remarks. V.B.B. acknowledges support from the DARPA DRINQS program and C.K. is supported by the DFG via the Emmy-Noether program under KA 3360/2-1.
\bibliography{afbib}

\onecolumngrid
\appendix

\section{Front-scaling for domain wall initial conditions at $\Delta=1$}

\begin{figure}[h]
\includegraphics[width=0.5\linewidth]{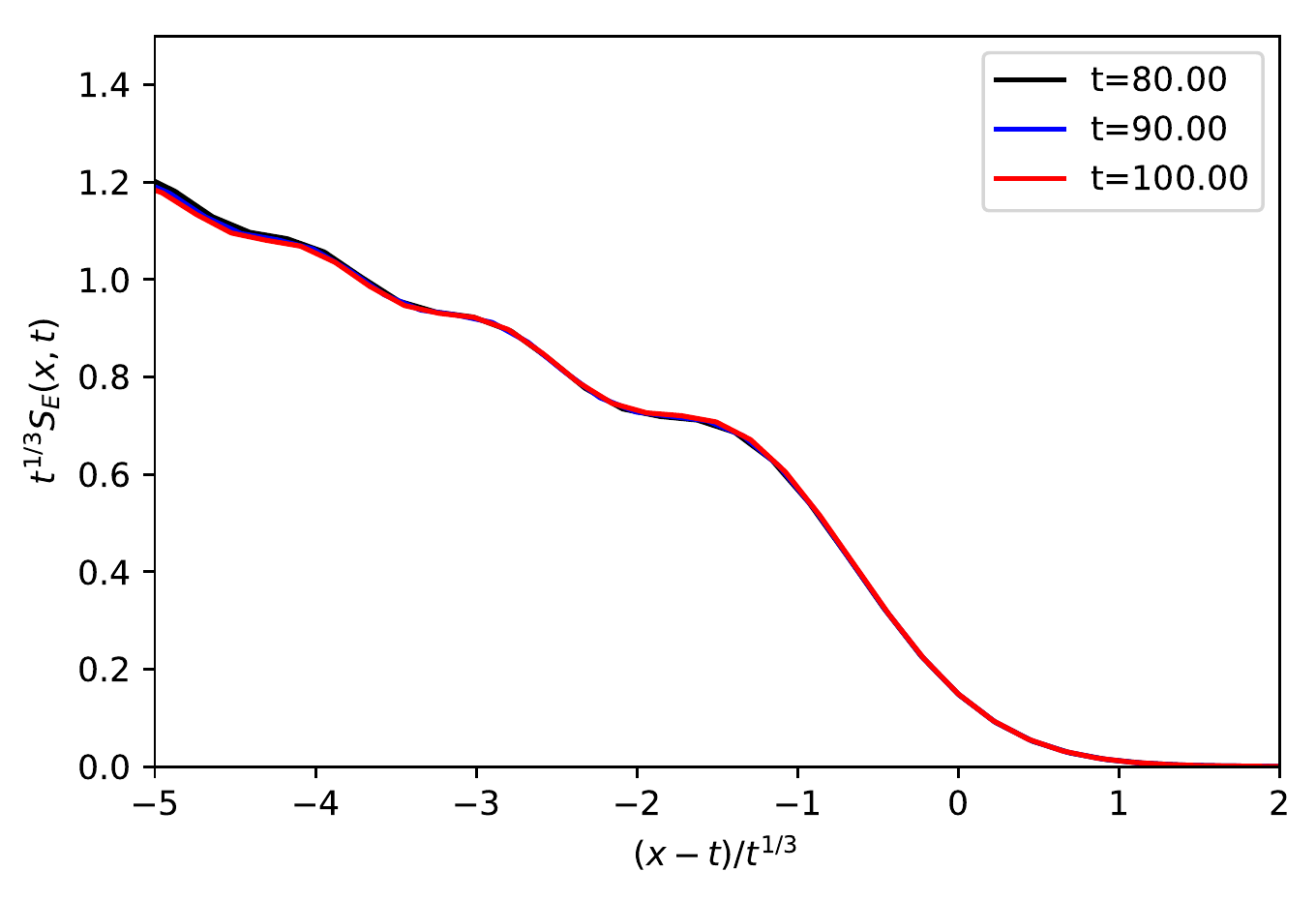}
\caption{Subdiffusive scaling collapse of fronts of single-cut entanglement entropy, obtained from tDMRG simulations of time-evolution from domain wall initial conditions in the XXZ chain, at anisotropy $\Delta=1$.}
\label{Fig3}
\end{figure}

\begin{figure}[h]
\includegraphics[width=\linewidth]{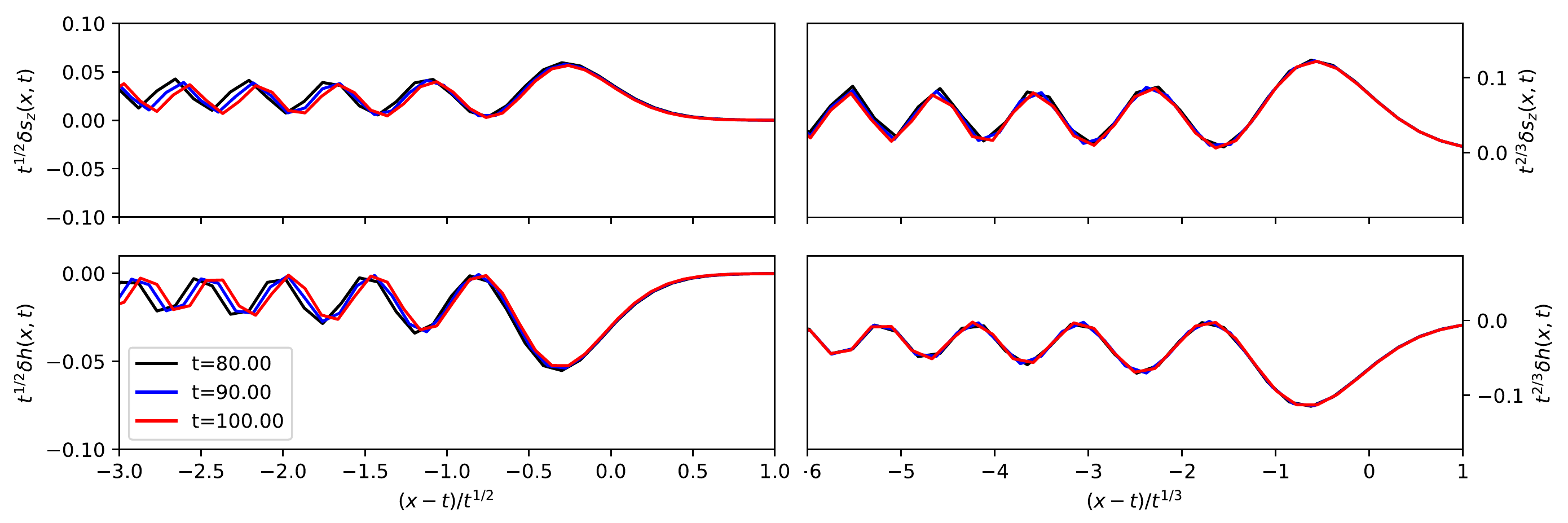}
\caption{Diffusive (\textit{left}) versus subdiffusive (\textit{right}) rescaling of fronts of spin (\textit{top}) and energy (\textit{bottom}), obtained from tDMRG predictions for time-evolution from domain wall initial conditions in the XXZ chain, at anisotropy $\Delta=1$.}
\label{Fig4}
\end{figure}

In this appendix, we include figures for the scaling of the ballistically spreading fronts of entanglement (Fig. \ref{Fig3}) and spin and energy (Fig. \ref{Fig4}), for time-evolution from domain wall initial conditions at the isotropic point, $\Delta=1$, of the XXZ chain.

\section{Exact spin, energy and entanglement entropy profiles for single spin-flip initial conditions}

\begin{figure}[t]
\includegraphics[width=0.6\linewidth]{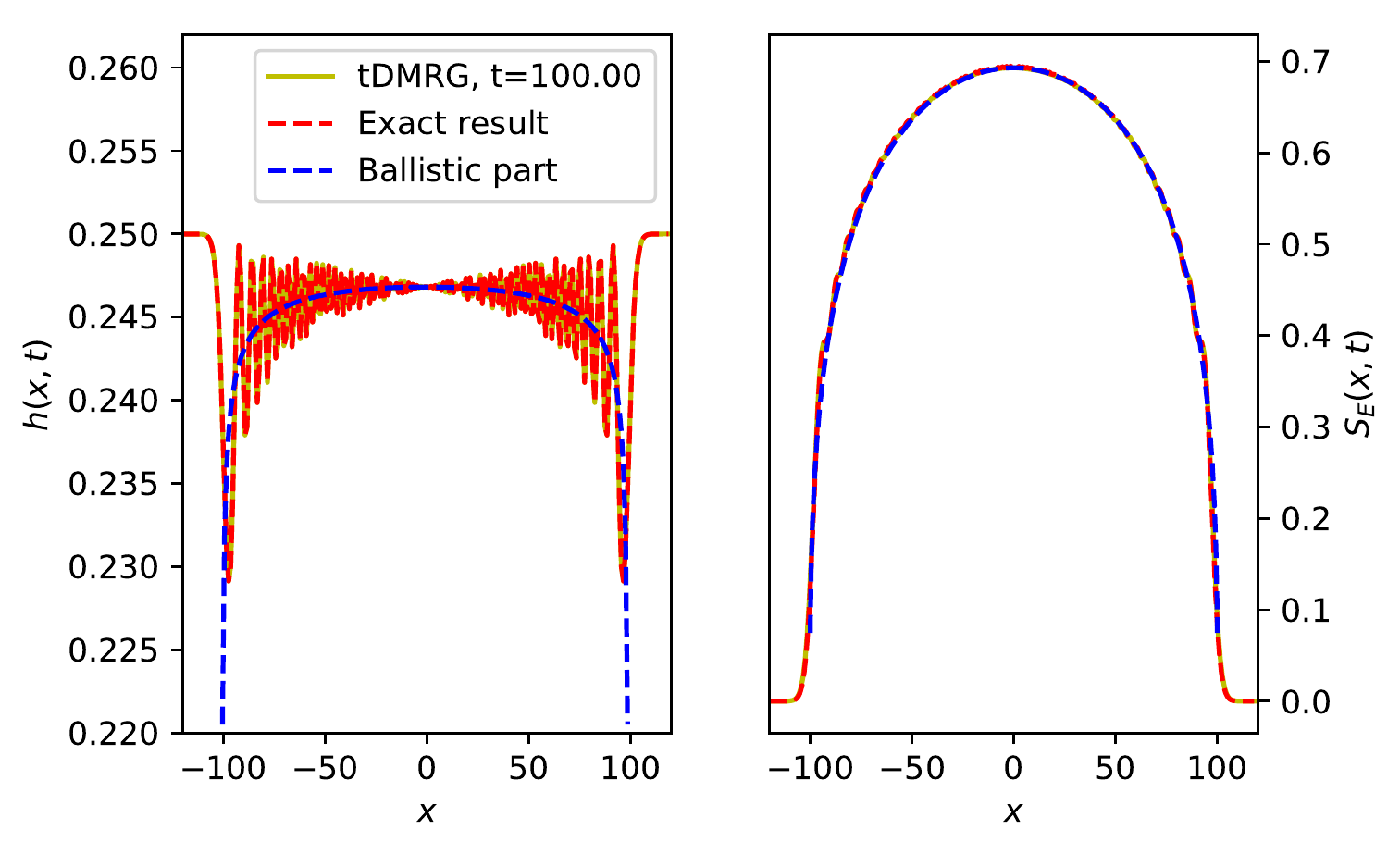}
\caption{\textit{Left:} Energy profile at $t=100$ obtained from tDMRG data for time-evolution from the initial condition Eq. \eqref{IC}, in the spin-$1/2$ XXZ chain with anisotropy $\Delta=1$. The exact result Eq. \eqref{exact} (red dash) is qualitatively very different from its ballistic part, Eq. \eqref{hydroflip} (blue dash) \textit{Right:} The same plot but for single-cut entanglement entropy. The three curves are almost indistinguishable.}
\label{Fig5}
\end{figure}

Following the analysis of Refs. \citen{Antal1999,Hunyadi2004}, the exact space-time profiles of spin, energy and entanglement for the initial condition
\begin{equation}
\label{IC}
|\psi\rangle =  \ket{\uparrow}^{\otimes L/2-1} \otimes  \ket{\downarrow} \otimes \ket{\uparrow}^{\otimes L/2},
\end{equation}
in the limit $L \to \infty$ are straightforwardly found to be
\begin{align}
\label{exact}
\nonumber \langle S^z_n \rangle (t) &= \frac{1}{2}-J_n^2(t), \\
\nonumber \langle h_n \rangle (t) &= \frac{J\Delta}{4} - \frac{J\Delta}{2}\left(J^2_n(t)+J_{n+1}^2(t)\right), \\
S_E(n,t) &= -\left(\sum_{m=-\infty}^n J_m^2(t)\right) \log{\left(\sum_{m=-\infty}^n J_m^2(t)\right)} - \left(\sum_{m=n+1}^\infty J_m^2(t)\right)\log{\left(\sum_{m=n+1}^\infty J_m^2(t)\right)},
\end{align}
where the $J_m(t)$ denote Bessel functions of the first kind. Passing to the ballistic scaling limit, these become
\begin{align}
\label{hydroflip} 
\nonumber s_z(x,t) &= \frac{1}{2} - \frac{1}{\pi}\frac{1}{t}\frac{1}{\sqrt{1-(x/t)^2}}, \\
\nonumber h(x,t) &= \frac{J\Delta}{4} - \frac{J\Delta}{\pi} \frac{1}{t} \frac{1}{\sqrt{1-(x/t)^2}}, \\
S_{E}(x,t) &= -\left(\frac{1}{2}-\frac{1}{\pi}\sin^{-1}\left(\frac{x}{t}\right)\right) \log{ \left(\frac{1}{2}-\frac{1}{\pi}\sin^{-1}\left(\frac{x}{t}\right)\right)} - \left(\frac{1}{2}+\frac{1}{\pi}\sin^{-1}\left(\frac{x}{t}\right)\right)\log{\left(\frac{1}{2}+\frac{1}{\pi}\sin^{-1}\left(\frac{x}{t}\right)\right)}.
\end{align}
The first two formulae can be obtained directly from considering the hydrodynamic time evolution of the initial state $\rho_k(x,0) = \delta(x)$. The comparison of the ballistic approximation with numerical results and the exact time-evolution is shown in Fig. \eqref{Fig5}. While the agreement for entanglement entropy is very good, the ballistic part of energy captures only the smoothed out profile, and the qualitative difference induced by third-order derivative terms is marked.

\begin{figure}[h]
\includegraphics[width=0.6\linewidth]{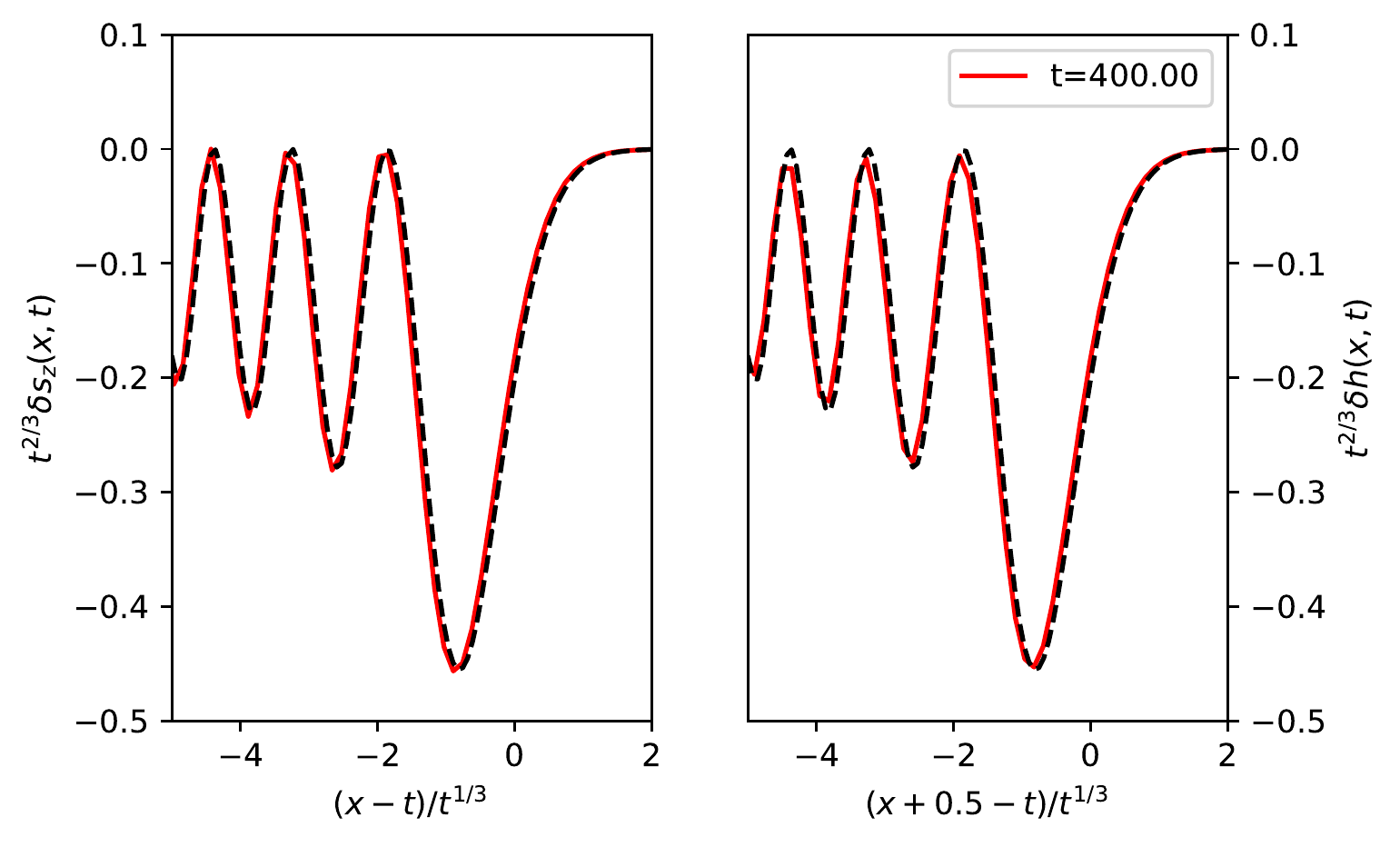}
\caption{Scaling collapse of fronts of spin and energy at $t=400$, obtained from tDMRG data for time-evolution from the initial condition Eq. \eqref{IC}, in the spin-$1/2$ XXZ chain with anisotropy $\Delta=1$. Exact asymptotic predictions (Eq. \eqref{frontanal}) are dashed.}
\label{Fig6}
\end{figure}

We now turn to the features that evolution from spin flips has in common with evolution from domain walls in the XXZ chain. For spin-flip initial conditions, the entanglement entropy does not appear to exhibit the $t^{-1/3}$ height characteristic of entanglement spreading from non-interacting domain walls\cite{Eisler2013}. However, for both spin-flips and domain walls, fronts of spin and energy exhibit the same $t^{-2/3}$ height. In the context of spin flips, this follows directly from Bessel function asymptotics in the transitional region\cite{Hunyadi2004}, which yield the front scaling
\begin{align}
\label{frontanal}
\delta s_z(x,t) \sim t^{-2/3} g_1\left(\frac{x-t}{t^{1/3}}\right), \quad  \delta h(x,t) \sim t^{-2/3}g_2\left(\frac{x+0.5-t}{t^{1/3}}\right),
\end{align}
where $\delta s_z(x,t) = \langle S^z \rangle (x,t)-1/2$, $\delta h (x,t) = \langle h \rangle(x,t)-J\Delta/4$ and
\begin{equation}
g_1(y) = -2^{2/3}[\mathrm{Ai}(2^{1/3}y)]^2, \quad g_2(y) = -J\Delta2^{2/3}[\mathrm{Ai}(2^{1/3}y)]^2.
\end{equation}
Here $\mathrm{Ai}(z)$ denotes the Airy function (the offset in the energy front is a transient due to the local Hamiltonian being a two-site operator, that is nevertheless necessary to obtain a good agreement with analytics). See Fig. \ref{Fig6} for a late-time comparison with these predictions at $\Delta=1$. 
\end{document}